\documentclass[conference]{IEEEtran}

\usepackage{cite}

\usepackage{amsmath,amssymb}
\usepackage{subfigure}
\usepackage[dvipdfmx]{graphicx}

\usepackage{stfloats}

\usepackage{bm}
\usepackage{comment}
\usepackage{color}

\usepackage{geometry}	
\usepackage{algcompatible}
\usepackage{algorithm}
\usepackage{algpseudocode}
\algrenewcommand\algorithmicindent{0em}
\makeatletter
\algnewcommand{\LineComment}[1]{\Statex \hskip\ALG@thistlm #1}

\makeatother
\usepackage{setspace}
\let\Algorithm\algorithm
\renewcommand\algorithm[1][]{\Algorithm[#1]\setstretch{1}}

\renewcommand{\baselinestretch}{0.948}
\IEEEoverridecommandlockouts

\usepackage[acronym,shortcuts]{glossaries}
\newacronym{CF-mMIMO}{CF-mMIMO}{cell-free massive multiple-input-multiple-output}
\newacronym{ISAC}{ISAC}{integrated sensing and communication}
\newacronym{TDD}{TDD}{time division duplex}
\newacronym{AP}{AP}{access point}
\newacronym{UE}{UE}{user equipment}
\newacronym{ULA}{ULA}{uniform linear array}
\newacronym{C-AP}{C-AP}{communication AP}
\newacronym{S-AP}{S-AP}{sensing AP}
\newacronym{i.i.d.}{i.i.d.}{independent and identically distributed}
\newacronym{CSI}{CSI}{channel state information}
\newacronym{AER}{AER}{activity error rate}
\newacronym{AMP}{AMP}{approximate message passing}
\newacronym{ANN}{ANN}{approximate nearest neighbors}
\newacronym{ASB}{ASB}{adaptively scaled belief}
\newacronym{AUD}{AUD}{active user detection}
\newacronym{AWGN}{AWGN}{additive white Gaussian noise}
\newacronym{AoA}{AoA}{angle of arrival}
\newacronym{AoD}{AoD}{angle of departure}
\newacronym{BAd-VAMP}{BAd-VAMP}{bilinear adaptive VAMP}
\newacronym{BBI}{BBI}{Bayesian bilinear inference}
\newacronym{BER}{BER}{bit error rate}
\newacronym{BG}{BG}{Bernoulli-Gaussian}
\newacronym{BMMSE}{BMMSE}{Bussgang minimum mean square error}
\newacronym{BiGAMP}{BiGAMP}{bilinear generalized approximate message passing}
\newacronym{BIP}{BIP}{bilinear inference problem}
\newacronym{GAMP}{GAMP}{generalized approximate message passing}
\newacronym{GF}{GF}{grant-free}
\newacronym{BiGaBP}{BiGaBP}{bilinear Gaussian belief propagation}
\newacronym{BP}{BP}{belief propagation}
\newacronym{BS}{BS}{base station}
\newacronym{BLER}{BLER}{block error rate}
\newacronym{CAP}{CAP}{central AP}
\newacronym{CCU}{CCU}{central computing unit}
\newacronym{CDF}{CDF}{cumulative distribution function}
\newacronym{CE}{CE}{channel estimation}
\newacronym{CFO}{CFO}{carrier frequency offset}
\newacronym{CIR}{CIR}{channel impulse response}
\newacronym{CLT}{CLT}{central limit theorem}
\newacronym{CP}{CP}{communication-prioritized}
\newacronym{CPU}{CPU}{central processing unit}
\newacronym{CRLB}{CRLB}{Cram\'{e}r–Rao bound}
\newacronym{CSIDCO}{CSIDCO}{complex SIDCO}
\newacronym{CSt-SBL}{CSt-SBL}{complex \textit{t}-distribution-based SBL}
\newacronym{DCC}{DCC}{dynamic cooperation clustering}
\newacronym{DFT}{DFT}{discrete Fourier transform}
\newacronym{DL}{DL}{deep learning}
\newacronym{DNN}{DNN}{deep neural network}
\newacronym{DSP}{DSP}{digital signal processor}
\newacronym{DoA}{DoA}{direction of arrival}
\newacronym{DoF}{DoF}{degrees of freedom}
\newacronym{DQ}{DQ}{De-quantization}
\newacronym{DU}{DU}{deep unfolding}
\newacronym{eMBB}{eMBB}{enhanced mobile broadband}
\newacronym{ECF}{ECF}{estimate-compress-forward}
\newacronym{EXIT}{EXIT}{extrinsic information transfer}
\newacronym{EM}{EM}{expectation-maximization}
\newacronym{EP}{EP}{expectation propagation}
\newacronym{FA}{FA}{false alarm}
\newacronym{FFT}{FFT}{fast Fourier transform}
\newacronym{FFNN}{FFNN}{feed-forward neural network}
\newacronym{FN}{FN}{factor node}
\newacronym{FG}{FG}{factor graph}
\newacronym{GaBP}{GaBP}{Gaussian belief propagation}
\newacronym{GM}{GM}{Gaussian-mixture}
\newacronym{IC}{IC}{interference cancellation}
\newacronym{IDD}{IDD}{iterative detection and decoding}
\newacronym{IFFT}{IFFT}{inverse fast Fourier transform}
\newacronym{ICI}{ICI}{inter-sub-carrier interference}
\newacronym{JACDE}{JACDE}{joint activity, channel and data estimation}
\newacronym{JACE}{JACE}{joint activity and channel estimation}
\newacronym{JCDE}{JCDE}{joint channel and data estimation}
\newacronym{JCCE}{JCCE}{joint channel and CFO estimation}
\newacronym{JCCDE}{JCCDE}{joint channel, CFO, and data estimation}
\newacronym{KL}{KL}{Kullback-Leibler}
\newacronym{LAMP}{LAMP}{learned AMP}
\newacronym{LSA}{LSA}{latent semantic analysis}
\newacronym{LoS}{LoS}{line-of-site}
\newacronym{LLR}{LLR}{log-likelihood ratio}
\newacronym{LMMSE}{LMMSE}{linear minimum mean square error}
\newacronym{LASSO}{LASSO}{least absolute shrinkage and selection operator}
\newacronym{MAC}{MAC}{multiple-access channel}
\newacronym{MAE}{MAE}{mean absolute error}
\newacronym{MAP}{MAP}{maximum \textit{a-posteriori}}
\newacronym{MASR}{MASR}{mainlobe-to-average-sidelobe ratio}
\newacronym{MCS}{MCS}{modulation and coding scheme}
\newacronym{MPDQ}{MPDQ}{message passing DQ}
\newacronym{MD}{MD}{miss-detection}
\newacronym{MF}{MF}{matched filter}
\newacronym{MFB}{MFB}{Matched filter bound}
\newacronym{MNS}{MNS}{minimum norm solution}
\newacronym{MI}{MI}{mutual information}
\newacronym{mMIMO}{mMIMO}{Massive multiple input multiple-output}
\newacronym{MIMO}{MIMO}{multiple-input multiple-output}
\newacronym{MIMO-OFDM}{MIMO-OFDM}{multiple-input multiple-output orthogonal frequency-division multiplexing}
\newacronym{MU-MIMO}{MU-MIMO}{multi-user multiple-input multiple-output}
\newacronym{MU-MIMO-OFDM}{MU-MIMO-OFDM}{multi-user multiple-input multiple-output orthogonal frequency-division multiplexing}
\newacronym{ML}{ML}{maximum likelihood}
\newacronym{MMSE}{MMSE}{minimum mean-square error}
\newacronym{MMV-AMP}{MMV-AMP}{multiple measurement vector approximate message passing}
\newacronym{MMV}{MMV}{multiple measurement vector}
\newacronym{MSE}{MSE}{mean square error}
\newacronym{MP}{MP}{message passing}
\newacronym{MPA}{MPA}{message passing algorithm}
\newacronym{MRC}{MRC}{maximal ratio combining}
\newacronym{MRT}{MRT}{maximum-ratio transmission}
\newacronym{MUD}{MUD}{multi-user detection}
\newacronym{mMTC}{mMTC}{massive machine type communications}
\newacronym{mmWave}{mmWave}{millimeter-wave}
\newacronym{NR}{NR}{new radio}
\newacronym{NMSE}{NMSE}{normalized mean square error}
\newacronym{OFDM}{OFDM}{orthogonal frequency-division multiplexing}
\newacronym{OLLA}{OLLA}{outer loop link adaptation}
\newacronym{PBI}{PBI}{parametric biliner inference}
\newacronym{PDA}{PDA}{probabilistic data association}
\newacronym{PDF}{PDF}{probability density function}
\newacronym{PE}{PE}{prediction error}
\newacronym{PMF}{PMF}{probability mass function}
\newacronym{PN}{PN}{phase noise}
\newacronym{PPP}{PPP}{Poisson point process}
\newacronym{PSK}{PSK}{phase-shift keying}
\newacronym{QoS}{QoS}{quality of service}
\newacronym{QP}{QP}{quadratic program}
\newacronym{QPSK}{QPSK}{quadrature PSK}
\newacronym{QAM}{QAM}{quadrature amplitude modulation}
\newacronym{RB}{RB}{resource block}
\newacronym{ReLU}{ReLU}{rectified linear unit}
\newacronym{RMSE}{RMSE}{root mean squared error}
\newacronym{RF}{RF}{radio frequency}
\newacronym{RX}{RX}{receive}
\newacronym{SAGE}{SAGE}{space-alternating generalized expectation-maximization}
\newacronym{SBL}{SBL}{sparse Bayesian learning}
\newacronym{SCA}{SCA}{successive convex approximation}
\newacronym{SD}{SD}{sphere decoding}
\newacronym{SE}{SE}{spectral efficiency}
\newacronym{SIDCO}{SIDCO}{sequential iterative decorrelation via convex optimization}
\newacronym{SGA}{SGA}{scalar Gaussian approximation}
\newacronym{SGD}{SGD}{stochastic gradient descent}
\newacronym{S-GAMP}{S-GAMP}{structured generalized approximate message passing}
\newacronym{SIC}{SIC}{soft interference cancellation}
\newacronym{SID}{SID}{self-iterative detection}
\newacronym{SIMO}{SIMO}{single-input multiple-output}
\newacronym{SINR}{SINR}{signal-to-interference-plus-noise ratio}
\newacronym{SNR}{SNR}{signal-to-noise power ratio}
\newacronym{Soft IC}{Soft IC}{soft interference cancellation}
\newacronym{SotA}{SotA}{state-of-the-art}
\newacronym{SP}{SP}{security-prioritized}
\newacronym{SPA}{SPA}{sum-product algorithm}
\newacronym{SSR}{SSR}{sparse signal recovery}
\newacronym{SVD}{SVD}{singular value decomposition}
\newacronym{TB}{TB}{transport block}
\newacronym{TDL}{TDL}{tapped delay line}
\newacronym{T-GaBP}{T-GaBP}{trainable GaBP}
\newacronym{T-GAMP}{T-GAMP}{trainable GAMP}
\newacronym{TX}{TX}{transmit}
\newacronym{URA}{URA}{uniform rectangular array}
\newacronym{URLLC}{URLLC}{ultra reliable low latency communications}
\newacronym{VAMP}{VAMP}{vector AMP}
\newacronym{VDB}{VDB}{vector database}
\newacronym{VGA}{VGA}{vector Gaussian approximation}
\newacronym{VN}{VN}{variable node}
\newacronym{VSS}{VSS}{vector similarity search}
\newacronym{ZF}{ZF}{zero-forcing}
\newacronym{flops}{flops}{floating point operations}
\newacronym{CS}{CS}{compressed sensing}
\newacronym{LIP}{LIP}{linear inference problem}
\newacronym{LS}{LS}{least square}
\newacronym{w.r.t.}{w.r.t.}{with respect to}

\begin{document}
%
\title{Secure Cell-Free Massive MIMO ISAC Systems: Joint AP Selection and Power Allocation\\ Against Eavesdropping}


\author{\IEEEauthorblockN{Ruiguang Wang$\,^*$, Takumi Takahashi$\,^*$, and Hideki Ochiai$\,^*$}
\IEEEauthorblockA{
$\,^*$ Graduate School of Engineering, The University of Osaka, 2-1 Yamada-oka, Suita, 565-0871, Japan\\ 
    Email: 
    $\,^*$\{r-wang@wcs., takahashi@, ochiai@\}comm.eng.osaka-u.ac.jp,\\ 
    \vspace{-5mm}
}}

\maketitle
\begin{abstract}
%
This paper investigates a \ac{CF-mMIMO} \ac{ISAC} system that addresses the critical challenge of information leakage to potential eavesdroppers located within sensing zones. 
A novel \ac{AP} selection strategy is proposed, which partitions the distributed \acp{AP} into two functional groups: \acp{C-AP}, dedicated exclusively to data transmission, and \acp{S-AP}, responsible for target detection and eavesdropper suppression. 
Closed-form expressions for the achievable communication rate, eavesdropping rate, and \ac{MASR} are derived to evaluate system performance.
Two complementary optimization problems are formulated using the \ac{SCA}: (i) maximizing user rates under security constraints and (ii) minimizing eavesdropping rates while satisfying \ac{QoS} requirements.
The proposed joint optimization framework determines the optimal \ac{AP} operational modes and power allocation across communication and sensing links.
Extensive numerical results validate the theoretical analysis and demonstrate significant performance gains, revealing inherent trade-offs among communication efficiency, sensing accuracy, and security.
These insights offer practical guidelines for designing secure \ac{CF-mMIMO} \ac{ISAC} systems. 
\end{abstract}

\vspace{1mm}

%
\IEEEpeerreviewmaketitle

\glsresetall

\section{Introduction}
\label{Chap1:intro}


Future 6G networks must deliver ultra-high communication rates while enabling precise environmental sensing for applications ranging from autonomous driving to digital twins~\cite{xu2022robust,yang2023novel}.
Meanwhile, \ac{CF-mMIMO} provides superior spectral efficiency and coverage uniformity~\cite{fang2021cell}.
The synergy between these technologies is compelling, as \ac{CF-mMIMO}'s distributed antenna arrays can simultaneously support high-throughput communication and high-precision radar sensing~\cite{liu2022integrated}.

However, \ac{CF-mMIMO}-based \ac{ISAC} systems face critical security vulnerabilities.
The unified waveform design allows sensing targets to potentially intercept confidential information~\cite{wei2022toward}.
Moreover, the reduced number of antennas per \ac{AP} weakens beamforming accuracy, which has conventionally served as a safeguard against eavesdropping.
While security improvements in traditional \ac{ISAC} systems have been thoroughly investigated~\cite{11124234}, the security of \ac{CF-mMIMO} \ac{ISAC} is still largely unexamined.

An optimal transmission strategy was proposed in~\cite{rivetti2024secure} that utilizes artificial noise to impair eavesdropper reception while minimizing the constrained Cramér-Rao bound via semidefinite relaxation, thus achieving a balance among sensing accuracy, communication quality, and security. 
 The study in~\cite{10605793} focuses on optimizing joint transmit beamforming among multiple \ac{ISAC} transmitters to enhance legitimate detection probability.
 It ensures communication quality and reduces the risks of data eavesdropping and sensing information interception by employing semidefinite relaxation, which has demonstrated global optimality.
Furthermore, \cite{11123583} introduces a joint optimization framework utilizing minorization-maximization and semidefinite programming to develop transmit beamforming and fronthaul compression, aimed at improving sensing performance while adhering to secure communication and capacity constraints. 
Unified waveforms, however, pose a risk of information leakage to sensing targets during sensing operations.

This paper investigates a \ac{CF-mMIMO} \ac{ISAC} system wherein distributed \acp{AP} simultaneously serve multiple \ac{UE} devices and sensing multiple zones.
We propose a novel joint \ac{AP} selection and power allocation method that assigns each \ac{AP} to either communication or sensing~\cite{10901970}, thereby reducing information leakage.
The primary contributions include: 
(i) the derivation of closed-form expressions for the communication rate, eavesdropping rate, and \ac{MASR}; (ii) the formulation of optimization problems aimed at maximizing communication rates or minimizing eavesdropping rates under sensing and power constraints; and (iii) validation of the proposed framework through comprehensive numerical results.

\textit{Notation}: Bold lowercase letters and uppercase letters denote vectors and matrices, respectively.
The superscript $\left(\cdot\right)^\dagger$ represents the conjugate transpose and $\bf{I}_N$ denotes the $N\times N$ identity matrix.
The statistical expectation operator is denoted by $\mathbb{E}\left[\cdot\right]$ and $\mathcal{CN}\left(\boldsymbol{\mu},\boldsymbol{\sigma}^2\right)$ represents a complex jointly Gaussian distribution with mean vector $\boldsymbol{\mu}$ and  covariance matrix $\boldsymbol{\sigma}^2$.
%

\section{System Model}
\label{Chap2:model}

\begin{figure}[H] 
    \centering 
    \includegraphics[scale=0.25]{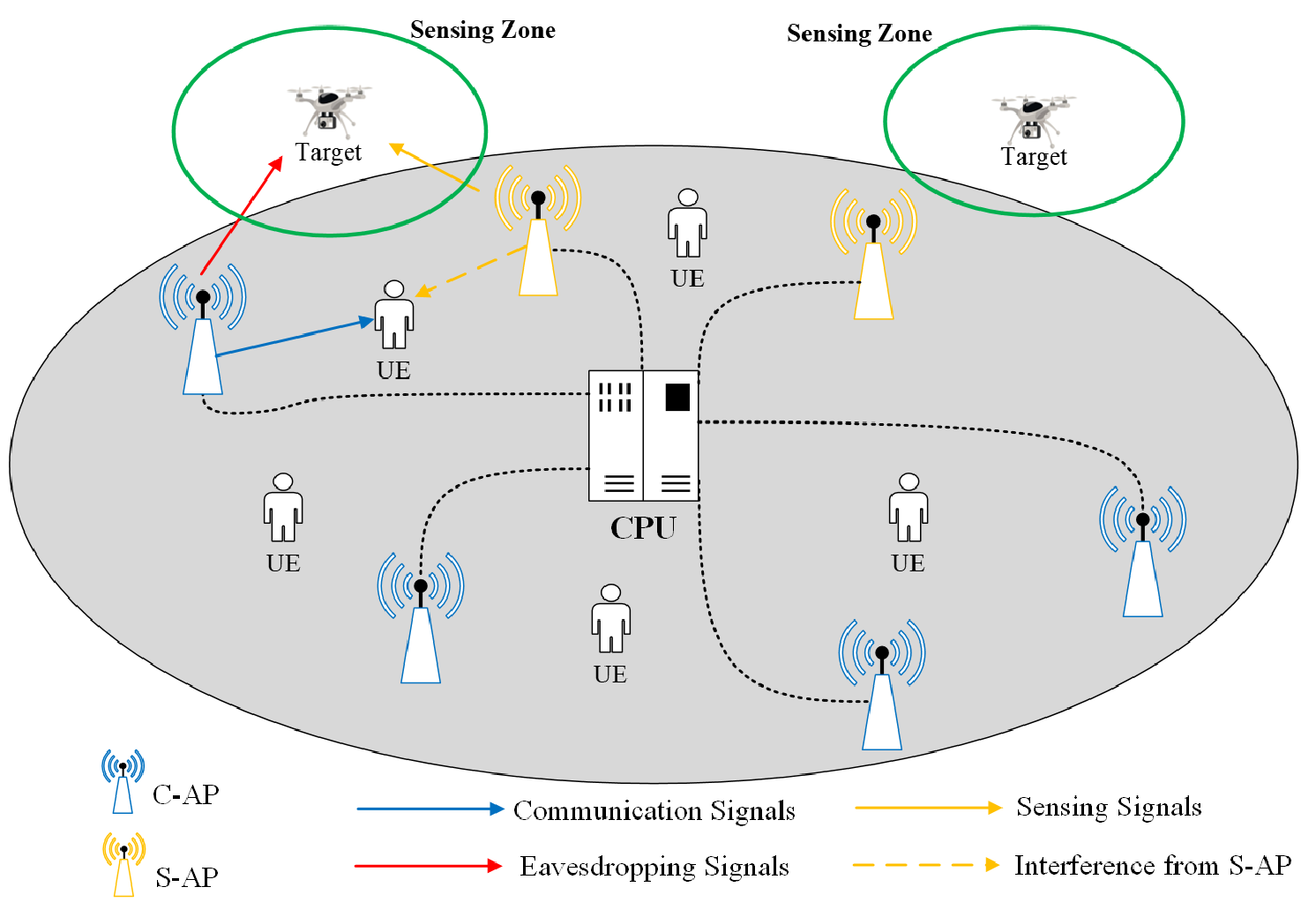} 
    \caption{Illustration of the \ac{CF-mMIMO} \ac{ISAC} system.} 
    \label{Fig.main2} 
\end{figure}

Consider a \ac{CF-mMIMO} \ac{ISAC} system illustrated in Fig.~\ref{Fig.main2}, where $M$ \acp{AP} simultaneously serve $K$ \acp{UE} while sensing $L$ zones.
Each \ac{AP} is equipped with $N$ antennas arranged in a \ac{ULA}, whereas both \acp{UE} and targets, which are regarded as potential eavesdroppers, located within the sensing zones are equipped with a single antenna.
To enhance security, we adopt an \ac{AP} selection strategy~\cite{10901970} that assigns \ac{AP} as either a \ac{C-AP} or a \ac{S-AP}, thereby preventing information leakage through separated signal transmission.

\subsection{Channel Model}

We assume quasi-static block fading channels that remain constant over coherence intervals of $\tau$ symbols, with $\tau_t$ symbols allocated for training.
The channel between \ac{AP} $m$ and \ac{UE} $k$ is expressed as $\bm{\mathbf{g}}_{mk}=\sqrt{\beta_{mk}}\bm{\mathbf{h}}_{mk}$, where $\beta_{mk}$ denotes the large-scale fading and $\bm{\mathbf{h}}_{mk}\sim\mathcal{CN}\left(\mathbf{0},\bm{\mathbf{I}}_N\right)$ represents the small-scale fading.
Using \ac{LMMSE} estimation~\cite{7827017}, the channel estimate is distributed as $\hat{\bm{\mathbf{g}}}_{mk}\sim\mathcal{CN}\left(\bm{0},\gamma_{mk}\bm{\mathbf{I}}_N\right)$, where $\gamma_{mk}\triangleq\frac{\tau_t\rho_t\beta^2_{mk}}{\tau_t\rho_t\beta_{mk}+1}$ and $\rho_t$ denotes the pilot transmit power.

For the sensing channel model, the \acp{AP} are assumed to possess prior knowledge of the sensing zones.
The direction from \ac{AP} $m$ to sensing zone $l$ is denoted by $\theta_{ml}$, representing the \ac{AoD} determined by their relative positions.
The sensing channel between \ac{AP} $m$ and sensing zone $l$ is modeled as $\bm{\mathbf{g}}_{ml}=\sqrt{\zeta_{ml}}\mathbf{a}_N\left(\theta_{ml}\right)$, where $\zeta_{ml}$ represents the distance-dependent path loss and $\mathbf{a}_N\left(\theta_{ml}\right)\in\mathbb{C}^{N\times1}$ is the array response vector whose $n$-th element is expressed as 
\begin{equation}
    \left[\mathbf{a}_N\left(\theta_{ml}\right)\right]_n = \mathrm{exp}\left[j\frac{2\pi d}{\lambda}\left(n-1\right)\sin\left(\theta_{ml}\right)\right],
\end{equation}
with $d$ and $\lambda$ representing the antenna spacing and carrier wavelength, respectively.

\subsection{Signal Structure and Transmission}

Let $\bm{\mathbf{x}}_m^c$ and $\bm{\mathbf{x}}_m^s$ denote the communication and sensing signals, respectively.
For later convenience, the transmitted signal at the $m$-th \ac{AP} is defined as 
\begin{equation}
\label{equ: signal_structure}
    \bm{\mathbf{x}}_m\triangleq\sqrt{a_m}\bm{\mathbf{x}}_m^c+\sqrt{1-a_m}\bm{\mathbf{x}}_m^s,
\end{equation}
where $a_m\in\left\{0,1\right\}$ is a binary variable indicating the operational mode of the $m$-th \ac{AP}.
During the downlink transmission phase, \acp{C-AP} transmit data signals to \acp{UE} employing \ac{MRT} precoding, while \acp{S-AP} simultaneously transmit sensing signals toward the sensing zones for target detection.
The transmitted signals from \acp{C-AP} and \acp{S-AP} are respectively formulated as follows:
\begin{equation}
    \bm{\mathbf{x}}_m^c=\sum\nolimits_{k=1}^K\sqrt{\rho\eta_{mk}^c}\hat{\mathbf{g}}_{mk}x_k^c,
\end{equation}
\begin{equation}
\label{equ: sensing signal}
    \bm{\mathbf{x}}_m^s=\sum\nolimits_{l=1}^L\sqrt{\rho\eta_{ml}^s}\mathbf{a}_N\left(\theta_{ml}\right)x_l^s,
\end{equation}
where $\rho$ denotes the normalized transmit power and $\hat{\mathbf{g}}_{mk}$ is the \ac{MRT} precoding vector.
The data symbols $x_k^c$ and sensing symbols $x_l^s$ are assumed to be independent and normalized, \textit{i.e.}, $\mathbb{E}\left\{|x_k^c|^2\right\}=\mathbb{E}\left\{|x_l^s|^2\right\}=1$.
Furthermore, $\eta_{mk}^c$ and $\eta_{ml}^s$ represent the power allocation coefficients, which must satisfy the per-\ac{AP} power constraint given by
\begin{equation}
    a_m\mathbb{E}\left\{\Vert\bm{\mathbf{x}}_m^c\Vert^2\right\}+\left(1-a_m\right)\mathbb{E}\left\{\Vert\mathbf{x}_m^s\Vert^2\right\}\leq\rho.
\end{equation}

Accordingly, the corresponding received signal at the $k$-th \ac{UE} can be expressed as
\begin{align}
    y_k&=\sum\nolimits_{m=1}^M\mathbf{g}^\dagger_{mk}\mathbf{x}_m+n_k \nonumber \\
    &=\sum\nolimits_{m=1}^M\sqrt{a_m\rho\eta_{mk}^c}\mathbf{g}^\dagger_{mk}\hat{\mathbf{g}}_{mk}x_k^c+n_k\\
    &\quad+\sum\nolimits_{k'\neq k}^K\sum\nolimits_{m=1}^M\sqrt{a_m\rho\eta_{mk'}^c}\mathbf{g}^\dagger_{mk}\hat{\mathbf{g}}_{mk'}x_{k'}^c\nonumber\\
    &\quad+\sum\nolimits_{m=1}^M\sum\nolimits_{l=1}^L\sqrt{\left(1-a_m\right)\rho\eta_{ml}^s}\mathbf{g}^\dagger_{mk}\mathbf{a}_N\left(\theta_{ml}\right)x_l^s, \nonumber
\end{align}
where $n_k$ denotes \ac{AWGN} with $n_k\sim\mathcal{CN}\left(0,1\right)$. 
Since the \acp{AP} lack instantaneous \ac{CSI}, the \ac{SINR} of the $k$-th \ac{UE} is derived using the \textit{use-and-then-forget} bounding technique~\cite{10118823}.
Accordingly, the \ac{SINR} for the $k$-th \ac{UE} can be expressed as
\begin{equation}
\label{equ: SINR_simplified}
\begin{split}
    &\mathrm{SINR_k}=\\
    &\frac{|\mathrm{DS}_k|^2}{\mathbb{E}\left\{|\mathrm{BU}_k|^2\right\}+\sum_{k'\neq k}^K\mathbb{E}\left\{|\mathrm{IUI}_{kk'}|^2\right\}+\mathbb{E}\left\{|\mathrm{IS}_k|^2\right\}+1},
\end{split}
\end{equation}
where
\begin{subequations}
\begin{align}
    &\mathrm{DS}_k\triangleq\mathbb{E}\left\{\sum\nolimits_{m=1}^M\sqrt{a_m\rho\eta_{mk}^c}\mathbf{g}^\dagger_{mk}\hat{\mathbf{g}}_{mk}\right\},\\
    &\mathrm{BU}_k\triangleq\left(\sum\nolimits_{m=1}^M\sqrt{a_m\rho\eta_{mk}^c}\mathbf{g}^\dagger_{mk}\hat{\mathbf{g}}_{mk}\right)-\mathrm{DS}_k,\\
    &\mathrm{IUI}_{kk'}\triangleq\sum\nolimits_{m=1}^M\sqrt{a_m\rho\eta_{mk'}^c}\mathbf{g}^\dagger_{mk}\hat{\mathbf{g}}_{mk'},\\
    &\mathrm{IS}_k\triangleq\sum\nolimits_{m=1}^M\sum\nolimits_{l=1}^L\sqrt{\left(1-a_m\right)\rho\eta_{ml}^s}\mathbf{g}^\dagger_{mk}\mathbf{a}_N\left(\theta_{ml}\right).
\end{align}
\end{subequations}
These terms respectively represent the strength of the desired signal, the beamforming uncertainty due to imperfect \ac{CSI}, inter-user interference, and interference from \acp{S-AP}.
The closed-form expression of \eqref{equ: SINR_simplified} is provided in \eqref{equ: SINR_K} at the top of the next page. 
\begin{figure*}[!t]
	\centering
	\begin{equation}
    \label{equ: SINR_K}
		\mathrm{SINR}_k=\frac{\left(N\sum_{m=1}^M\sqrt{a_m\rho\eta_{mk}^c}\gamma_{mk}\right)^2}{N\sum_{k'=1}^K\sum_{m=1}^Ma_m\rho\eta_{mk'}^c\beta_{mk}\gamma_{mk'}+N\sum_{m=1}^M\sum_{l=1}^L\left(1-a_m\right)\rho\eta_{ml}^s\beta_{mk}+1}
	\end{equation}
    \vspace{-4mm}
\end{figure*}

\subsection{Eavesdropping and Sensing Operation}

In this subsection, we focus on the sensing operation, which serves the dual purpose of target detection and physical-layer security enhancement.
Concurrent with the downlink data transmission from \acp{C-AP}, \acp{S-AP} transmit dedicated sensing signals, as given in \eqref{equ: sensing signal}, toward the sensing zones.
To quantify detection effectiveness, the sensing performance is evaluated using the \ac{MASR} at the sensing-zone locations, defined as 
\begin{equation}
    \mathrm{MASR}_l\triangleq\frac{P_{\mathrm{DS}}^{\mathrm{Sen}}\left(\theta_{ml}\right)}{P^{\mathrm{Com}}\left(\theta_{ml}\right)+P_{\mathrm{DST}}^{\mathrm{Sen}}\left(\theta_{ml}\right)}, 
\end{equation}
where
\begin{subequations}
\begin{align}
    &P^{\mathrm{Com}}\left(\theta_{ml}\right)\triangleq\mathbb{E}\left\{|\sum\nolimits_{m=1}^M\sqrt{a_m}\mathbf{g}^\dagger_{ml}\bm{\mathbf{x}}_m^c|^2\right\},\\
    &P_{\mathrm{DS}}^{\mathrm{Sen}}\left(\theta_{ml}\right)\triangleq\nonumber\\
    &\mathbb{E}\left\{|\sum\nolimits_{m=1}^M\sqrt{\left(1-a_m\right)\rho\eta_{ml}^s}\mathbf{g}^\dagger_{ml}\mathbf{a}_N\left(\theta_{ml}\right)x_l^s|^2\right\},\\
    &P_{\mathrm{DST}}^{\mathrm{Sen}}\left(\theta_{ml}\right)\triangleq\mathbb{E}\left\{|\sum\nolimits_{l'=1}^L\sum\nolimits_{m=1}^M\sqrt{\left(1-a_m\right)\rho\eta_{ml'}^s}\right.\nonumber\\
    &\times\left.\mathbf{g}^\dagger_{ml}\mathbf{a}_N\left(\theta_{ml'}\right)x_{l'}^s|^2\right\}.
    \end{align}
\end{subequations}
These respectively represent the communication power pattern, the desired sensing power pattern, and the distortion power pattern.
Notably, the communication power pattern introduces distortion to the sensing operation, thereby degrading detection accuracy.
The detailed expression of $\mathrm{MASR}_l$ is provided in \eqref{equ: MASR_l} at the top of this page.
\begin{figure*}[!t]
	\centering
	\begin{equation}
    \label{equ: MASR_l}
		\mathrm{MASR}_l=\frac{N^2\sum_{m=1}^M\left(1-a_m\right)\eta_{ml}^s\zeta_{ml}}{N\sum_{k=1}^K\sum_{m=1}^Ma_m\eta_{mk}^c\zeta_{ml}\gamma_{mk}+\sum_{m=1}^M\sum_{l'\neq l}^L\left(1-a_m\right)\eta_{ml'}^s\zeta_{ml}|\mathbf{a}_N^\dagger\left(\theta_{ml}\right)\mathbf{a}_N\left(\theta_{ml'}\right)|^2}
	\end{equation}
    \hrule
    \vspace{-2mm}
\end{figure*}

Meanwhile, multiple potential eavesdroppers are assumed to be located within the sensing zones, attempting to intercept confidential data transmitted by \acp{C-AP}. 
The signal received by an eavesdropper at the $l$-th sensing zone is expressed as
\begin{align}
    y_l&=\sum\nolimits_{k=1}^K\sum\nolimits_{m=1}^M\sqrt{a_m\rho\eta_{mk}^c}\mathbf{g}^\dagger_{ml}\hat{\mathbf{g}}_{mk}x_k^c+\nonumber\\
    &\sum\nolimits_{m=1}^M\sum\nolimits_{l'=1}^L\sqrt{\left(1-a_m\right)\rho\eta_{ml'}^s}\mathbf{g}^\dagger_{ml}\mathbf{a}_N\left(\theta_{ml'}\right)x_{l'}^s+n_l,
\end{align}
where the first, second, and third terms correspond to the communication signals transmitted by \acp{C-AP}, the sensing signals transmitted by \acp{S-AP}, and \ac{AWGN}, respectively. 
From the eavesdroppers' perspective, the communication signals act as desired signals, whereas the sensing signals function as intentionally introduced interference or artificial noise. 
Accordingly, the resulting \ac{SINR} at the eavesdropper can be expressed as 
\begin{equation}
    \mathrm{SINR_l}=\frac{\mathbb{E}\left\{|\mathrm{DS}_l|^2\right\}}{\mathbb{E}\left\{|\mathrm{IS}_l|^2\right\}+1},
\end{equation}
where
\begin{equation}
    \mathrm{DS}_l\triangleq\sum\nolimits_{k=1}^K\sum\nolimits_{m=1}^M\sqrt{a_m\rho\eta_{mk}^c}\mathbf{g}^\dagger_{ml}\hat{\mathbf{g}}_{mk}
\end{equation}
and
\begin{equation}
    \mathrm{IS}_l\triangleq\sum\nolimits_{m=1}^M\sum\nolimits_{l'=1}^L\sqrt{\left(1-a_m\right)\rho\eta_{ml'}^s}\mathbf{g}^\dagger_{ml}\mathbf{a}_N\left(\theta_{ml'}\right)
\end{equation}
represent the desired eavesdropping signal and the interference induced by sensing signals, respectively.
The corresponding closed-form \ac{SINR} expression is presented in \eqref{equ: SINR_l} at the top of the next page. 
\begin{figure*}[!t]
	\centering
	\begin{equation}
    \label{equ: SINR_l}
		\mathrm{SINR}_l=\frac{N\sum_{k=1}^K\sum_{m=1}^Ma_m\rho\eta_{mk}^c\zeta_{ml}\gamma_{mk}}{\sum_{m=1}^M\sum_{l'=1}^L\left(1-a_m\right)\rho\eta_{ml'}^s\zeta_{ml}|\mathbf{a}_N^\dagger\left(\theta_{ml}\right)\mathbf{a}_N\left(\theta_{ml'}\right)|^2+1}
	\end{equation}
    \hrule
    \vspace{-2mm}
\end{figure*}
While sensing signals act as interference to degrade eavesdropper reception, the overall security performance strongly depends on resource allocation.
The key challenge, therefore, lies in the joint optimization of \ac{AP} operational modes $a_m$ and power allocation across communication and sensing links, as formulated in the following section.

\section{Joint Power Allocation}
\label{Chap3:opt_method}


We formulate a resource allocation problem that encapsulates the inherent trade-offs among communication, sensing, and security.
Increasing power allocation to \acp{C-AP} improves communication performance but raises the risk of eavesdropping.
In contrast, allocating power to  \acp{S-AP} enhances jamming and sensing performance at the expense of reduced data rates.
To tackle this trade-off, we jointly optimize \ac{AP} operational modes and power allocation using two complementary approaches: (i) maximizing the communication rate while adhering to security constraints, or (ii) minimizing the eavesdropping rate while satisfying \ac{QoS} constraints. 
In both cases, a minimum sensing \ac{MASR} threshold is imposed for all $L$ sensing zones.

By invoking \eqref{equ: signal_structure}, the total transmit power at each \ac{AP} is constrained by a maximum power budget, given by
\begin{equation}
    N\sum\nolimits_{k=1}^Ka_m\eta_{mk}^c\gamma_{mk}+N\sum\nolimits_{l=1}^L\left(1-a_m\right)\eta_{ml}^s\leq1.
\end{equation}

\subsection{Communication-Prioritized Optimization Method}

This subsection presents the first optimization problem, which prioritizes communication performance by maximizing the worst-case \ac{UE} \ac{SINR} to ensure max-min fairness. 
For notational convenience, we first define $\mathbf{a}\triangleq\left\{a_1,\dots,a_M\right\},\bm{\eta}^c\triangleq\left\{\eta_{m1}^c,\dots,\eta_{mK}^c\right\}$, and $\bm{\eta}^s\triangleq\left\{\eta_{m1}^s,\dots,\eta_{mL}^s\right\}$ for all $m$. 
The optimization problem is then formulated as
\begin{subequations}
\begin{align}
        \bm{\left(\mathrm{P1}\right):}&\max_{\bm{a},\bm{\eta}^c,\bm{\eta}^s}\min_k\;\mathrm{SINR}_k\left(\bm{a},\bm{\eta}^c,\bm{\eta}^s\right)\nonumber\\
        &\mathrm{s.t.}\;\mathrm{SINR}_l\left(\bm{a},\bm{\eta}^c,\bm{\eta}^s\right)\leq\nu,\forall l,\label{P1:const1}\\
        &\quad\ \ \mathrm{MASR}_l\left(\bm{a},\bm{\eta}^c,\bm{\eta}^s\right)\geq\kappa,\forall l,\label{P1:const2}\\
        &\quad\ \ N\sum\nolimits_{k=1}^K\eta_{mk}^c\gamma_{mk}\leq a_m,\forall m,\label{P1:const3}\\
        &\quad\ \ N\sum\nolimits_{l=1}^L\eta_{ml}^s\leq1-a_m,\forall m,\label{P1:const4}\\
        &\quad\ \ \ a_m\in\left\{0,1\right\},\forall m,\label{P1:const5}&
\end{align}
\end{subequations}
where \eqref{P1:const1} constrains the eavesdropper \ac{SINR}, \eqref{P1:const2} ensures the minimum sensing \ac{MASR}, \eqref{P1:const3} and \eqref{P1:const4} impose the per-\ac{AP} power constraints for \acp{C-AP} and \acp{S-AP}, and \eqref{P1:const5} enforces binary \ac{AP} selection.

First, we introduce a slack variable $t\triangleq\min_{k}\mathrm{SINR}_k$.
To address the integer binary constraint, the binary variables are relaxed. 
The constraint $a_m\in\left\{0,1\right\}$ can be equivalently expressed as $a_m=a_m^2$. 
Meanwhile, the inequality $a_m^2\leq a_m$ always holds for $a_m\in\left[0,1\right]$.
Following~\cite{6816086}, a penalty term with parameter $\lambda$ is added to drive $a_m$ toward binary values, and $a_m^2$ is substituted in the power constraints for computational efficiency.
The resulting non-convexity is then handled using the \ac{SCA} method.
To reformulate the optimization problem, we apply the following inequality:
\begin{equation}
\label{ieq: apr1}
    x^2\geq x_0\left(2x-x_0\right),
\end{equation}
where $x$ and $x_0$ are arbitrary real numbers.

Following the above transformations, the optimization problem is reformulated as 
\begin{subequations}
\begin{align}
        \bm{\left(\mathrm{P2}\right):}&\max_{\bm{a},\bm{\eta}^c,\bm{\eta}^s,t}\;t-\lambda\sum\nolimits_{m=1}^Ma_m-a_m^{\left(n\right)}\left(2a_m-a_m^{\left(n\right)}\right)\nonumber\\
        \!\!\!\!\!\!\!\!\!\!\!\!&\!\!\!\!\!\!\!\!\!\!\!\!\mathrm{s.t.}\;\mathrm{SINR}_k\left(\bm{a},\bm{\eta}^c,\bm{\eta}^s\right)\geq t,\forall m,\label{P2:const1}\\
        \!\!\!\!\!\!\!\!\!\!\!\!&\!\!\!\!\!\!\!\!\!\!\!\!\quad\ \ \mathrm{SINR}_l\left(\bm{a},\bm{\eta}^c,\bm{\eta}^s\right)\leq\nu,\forall l,\label{P2:const2}\\
        \!\!\!\!\!\!\!\!\!\!\!\!&\!\!\!\!\!\!\!\!\!\!\!\!\quad\ \ \mathrm{MASR}_l\left(\bm{a},\bm{\eta}^c,\bm{\eta}^s\right)\geq\kappa,\forall l,\label{P2:const3}\\
        \!\!\!\!\!\!\!\!\!\!\!\!&\!\!\!\!\!\!\!\!\!\!\!\!\quad\ \ N\sum\nolimits_{k=1}^K\eta_{mk}^c\gamma_{mk}\leq a_m^{\left(n\right)}\left(2a_m-a_m^{\left(n\right)}\right),\forall m,\label{P2:const4}\\
        \!\!\!\!\!\!\!\!\!\!\!\!&\!\!\!\!\!\!\!\!\!\!\!\!\quad\ \ N\sum\nolimits_{l=1}^L\eta_{ml}^s\leq\left(1-a_m^2\right),\forall m,\label{P2:const5}\\
        \!\!\!\!\!\!\!\!\!\!\!\!&\!\!\!\!\!\!\!\!\!\!\!\!\quad\ \ 
        0\leq a_m\leq 1,\forall m.\label{P2:const6}
\end{align}
\end{subequations}
where $\left(\cdot\right)^{\left(n\right)}$ denotes the feasible solution in the $n$-th iteration. 

Having relaxed the binary variables and power constraints, the main challenge arises from the non-convexity of the constraints \eqref{P2:const1}--\eqref{P2:const3}. 
This issue is addressed using \ac{SCA}.
Specifically, constraint \eqref{P2:const1} is reformulated as 
\begin{align}
        &\frac{\left(N\sum_{m=1}^M\sqrt{a_m\rho\eta_{mk}^c}\gamma_{mk}\right)^2}{t}\geq\rho\sum\nolimits_{m=1}^Ma_m\mu_{mk}\nonumber\\
        &+N\sum\nolimits_{m=1}^M\sum\nolimits_{l=1}^L\rho\eta_{ml}^s\beta_{mk}+1,\label{eq: constraint1}
\end{align}
where
\begin{equation}
\label{eq: ease_discrip1}
    \mu_{mk}\triangleq N\sum\nolimits_{k'=1}^K\eta_{mk'}^c\beta_{mk}\gamma_{mk'}-N\sum\nolimits_{l=1}^L\eta_{ml}^s\beta_{mk}.
\end{equation}
Then, \eqref{eq: constraint1} can be reformulated as
\begin{align}
        &\frac{\left(2N\sum_{m=1}^M\sqrt{a_m\eta_{mk}^c}\gamma_{mk}\right)^2}{t}+\sum\nolimits_{m=1}^M\left(a_m-\mu_{mk}\right)^2\geq\nonumber\\        &\sum\nolimits_{m=1}^M\left(a_m+\mu_{mk}\right)^2+4N\sum\nolimits_{m=1}^M\sum\nolimits_{l=1}^L\eta_{ml}^s\beta_{mk}+4{/}\rho.\label{ieq: cst_SINR}
\end{align}
To obtain a concave lower bound for the left-hand side of the above inequality, we apply the following relationship: 
\begin{equation}
\label{ieq: apr2}
    \frac{x^2}{y}\geq\frac{x_0}{y_0}\left(2x-\frac{x_0}{y_0}y\right), y>0,
\end{equation}
where $y, y_0$ denote arbitrary real numbers.

Using \eqref{ieq: apr1} and \eqref{ieq: apr2}, \eqref{ieq: cst_SINR} can be reformulated as
\begin{align}
        &q_k^{\left(n\right)}\left(4N\sum\nolimits_{m=1}^M\sqrt{a_m\eta_{mk}^c}\gamma_{mk}-q_k^{\left(n\right)}t\right)+\nonumber\\
        &\sum\nolimits_{m=1}^M\left(a_m^{\left(n\right)}-\mu_{mk}^{\left(n\right)}\right)\left(2\left(a_m-\mu_{mk}\right)-\left(a_m^{\left(n\right)}-\mu_{mk}^{\left(n\right)}\right)\right)\geq\nonumber\\
        &\sum\nolimits_{m=1}^M\left(a_m+\mu_{mk}\right)^2+4N\sum\nolimits_{m=1}^M\sum\nolimits_{l=1}^L\eta_{ml}^s\beta_{mk}+4{/}\rho,\label{P3: constraint1}
\end{align}
where
\begin{equation}
    q_k^{\left(n\right)}\triangleq\frac{2N\sum_{m=1}^M\sqrt{a_m^{\left(n\right)}\left(\eta_{mk}^c\right)^{\left(n\right)}}\gamma_{mk}}{t^{\left(n\right)}}.
\end{equation}
%

Following the same procedure and invoking \eqref{ieq: apr1}, we recast \eqref{P2:const2} and \eqref{P2:const3} as 
\begin{align}
        &\sum\nolimits_{m=1}^M\left(a_m^{\left(n\right)}-\varrho_{ml}^{\left(n\right)}\right)\left(2\left(a_m-\varrho_{ml}\right)-\left(a_m^{\left(n\right)}-\varrho_{ml}^{\left(n\right)}\right)\right)\nonumber\\
        &+4\nu\sum\nolimits_{m=1}^M\sum\nolimits_{l'=1}^L\eta_{ml'}^s\zeta_{ml}|\mathbf{a}_N^\dagger\left(\theta_{ml}\right)\mathbf{a}_N\left(\theta_{ml'}\right)|^2+\frac{4\nu}{\rho}\nonumber\\
        &\geq\sum\nolimits_{m=1}^M\left(a_m+\varrho_{ml}\right)^2,\label{P3: constraint2}
\end{align}
and
\begin{align}
        &4N^2\sum\nolimits_{m=1}^M\eta_{ml}^s\zeta_{ml}+\sum\nolimits_{m=1}^M\left(a_m^{\left(n\right)}-\omega_{ml}^{\left(n\right)}\right)\nonumber\\
        &\times\left(2\left(a_m-\omega_{ml}\right)-\left(a_m^{\left(n\right)}-\omega_{ml}^{\left(n\right)}\right)\right)\geq\nonumber\\
        &4\kappa\sum\nolimits_{l'\neq l}^L\eta_{ml'}^s\zeta_{ml}|\mathbf{a}_N^\dagger\left(\theta_{ml}\right)\mathbf{a}_N\left(\theta_{ml'}\right)|^2\nonumber\\
        &+\sum\nolimits_{m=1}^M\left(a_m+\omega_{ml}\right)^2,\label{P3: constraint3}
\end{align}
where
\begin{align}
        \varrho_{ml}&\triangleq N\sum\nolimits_{k=1}^K\eta_{mk}^c\zeta_{ml}\gamma_{mk}\nonumber\\
        &+\nu\sum\nolimits_{l'=1}^L\eta_{ml'}^s\zeta_{ml}|\mathbf{a}_N^\dagger\left(\theta_{ml}\right)\mathbf{a}_N\left(\theta_{ml'}\right)|^2,\\
        \omega_{ml}&\triangleq N^2\eta_{ml}^s\zeta_{ml}+\kappa\sum\nolimits_{k=1}^KN\eta_{mk}^c\zeta_{ml}\gamma_{mk}\nonumber\\
        &-\kappa\sum\nolimits_{l'\neq l}^L\eta_{ml'}^s\zeta_{ml}|\mathbf{a}_N^\dagger\left(\theta_{ml}\right)\mathbf{a}_N\left(\theta_{ml'}\right)|^2.\label{eq: omega}
\end{align}
%

Substituting the non-convex constraints with their convex approximations transforms the problem into a standard convex program, given by 
\begin{align}
        \bm{\left(\mathrm{P3}\right):}&\max_{\bm{a},\bm{\eta}^c,\bm{\eta}^s,t}\;t-  \lambda\sum\nolimits_{m=1}^Ma_m-a_m^{\left(n\right)}\left(2a_m-a_m^{\left(n\right)}\right)\nonumber\\
        &\mathrm{s.t.}\;\eqref{P3: constraint1}, \eqref{P3: constraint2}, \eqref{P3: constraint3}, \eqref{P2:const4}, \eqref{P2:const5}, \eqref{P2:const6}.\nonumber
\end{align}

This reformulated problem adheres to convex optimization theory and can be efficiently solved using standard solvers such as CVX. 
The iterative procedure starts from a randomly generated feasible point $\bf{\tilde{x}}\triangleq\left\{\bf{a},\boldsymbol{\eta}^c,\boldsymbol{\eta}^s\right\}$ and updates the solution iteratively.
The obtained solution $\bf{\tilde{x}}^*$ is then used as the initial point for the next iteration, and the process continues until convergence.

\subsection{Security-Prioritized Optimization Method}

The second optimization strategy prioritizes security by minimizing the maximum eavesdropper \ac{SINR} while maintaining minimum \ac{SINR} thresholds for \acp{UE}.
This formulation inverts the objective-constraint relationship of Problem $\left(\bf{P1}\right)$, where communication performance is treated as constraint rather than an objective.
By applying the same relaxation technique as in Problem $\left(\bf{P1}\right)$, the optimization problem is formulated as
\begin{subequations}
\begin{align}
        \bm{\left(\mathrm{P4}\right):}&\min_{\bm{a},\bm{\eta}^c,\bm{\eta}^s,t}\;t+\lambda\sum\nolimits_{m=1}^Ma_m-a_m^{\left(n\right)}\left(2a_m-a_m^{\left(n\right)}\right)\nonumber\\
        &\mathrm{s.t.}\;\mathrm{SINR}_l\left(\bm{a},\bm{\eta}^c,\bm{\eta}^s\right)\leq t,\forall m,\label{P4: constraint1}\\
        &\quad\ \ \ \mathrm{SINR}_k\left(\bm{a},\bm{\eta}^c,\bm{\eta}^s\right)\geq\varsigma,\forall l,\label{P4: constraint2}\\
        &\quad\quad\eqref{P2:const3}, \eqref{P2:const4}, \eqref{P2:const5}, \eqref{P2:const6}.\nonumber
\end{align}
\end{subequations}
The sensing performance constraint is treated in the same manner as in the communication-prioritized optimization method; hence, the detailed derivation is omitted for brevity. 
We now address the non-convex constraints \eqref{P4: constraint1} and \eqref{P4: constraint2}.
Specifically, constraint \eqref{P4: constraint1} is reformulated as 
%
\begin{equation}
\begin{split}
    &\sum\nolimits_{m=1}^M\left(t+\varepsilon_{ml}\right)^2+\sum\nolimits_{m=1}^M\left(a_m-\delta_{ml}\right)^2+4t{/}\rho\\
    &\geq\sum\nolimits_{m=1}^M\left(t-\varepsilon_{ml}\right)^2+\sum\nolimits_{m=1}^M\left(a_m+\delta_{ml}\right)^2\\
    &+4f_{ml}\left(t,a_m,\varepsilon_{ml}\right)
\end{split}
\end{equation}
where
\begin{subequations}
\begin{align}
    &\delta_{ml}=N\sum\nolimits_{k=1}^K\eta_{mk}^c\zeta_{ml}\gamma_{mk},\\&\varepsilon_{ml}=\sum\nolimits_{l'=1}^L\eta_{ml'}^s\zeta_{ml}|\mathbf{a}_N^\dagger\left(\theta_{ml}\right)\mathbf{a}_N\left(\theta_{ml'}\right)|^2,\\
    &f_{ml}\left(t,a_m,\varepsilon_{ml}\right)=t\sum\nolimits_{m=1}^Ma_m\varepsilon_{ml}.
\end{align}
\end{subequations}
For the non-convex function $f_{ml}\left(t,a_m,\varepsilon_{ml}\right)$, we form a convex approximation by applying a first-order Taylor expansion as 
\begin{equation}
\label{apr: taylor}
\begin{split}
        f_{ml}^{\left(n\right)}\left(t,a_m,\varepsilon_{ml}\right)&=\sum\nolimits_{m=1}^M\left(a_m^{\left(n\right)}\varepsilon_{ml}^{\left(n\right)}t+a_m^{\left(n\right)}t^{\left(n\right)}\varepsilon_{ml}\right.\\
        &\left.+\varepsilon_{ml}^{\left(n\right)}t^{\left(n\right)}a_m-2a_m^{\left(n\right)}\varepsilon_{ml}^{\left(n\right)}t^{\left(n\right)}\right).
\end{split}
\end{equation}
%
By invoking \eqref{ieq: apr1} and \eqref{apr: taylor}, we obtain a concave lower bound of the left-hand side, as follows:
\begin{align}
    &\sum\nolimits_{m=1}^M\left(t^{\left(n\right)}+\varepsilon_{ml}^{\left(n\right)}\right)\left(2\left(t+\varepsilon_{ml}\right)-\left(t^{\left(n\right)}+\varepsilon_{ml}^{\left(n\right)}\right)\right)\nonumber\\
    &+\sum\nolimits_{m=1}^M\left(a_m^{\left(n\right)}-\delta_{ml}^{\left(n\right)}\right)\left(2\left(a_m-\delta_{ml}\right)-\left(a_m^{\left(n\right)}-\delta_{ml}^{\left(n\right)}\right)\right)\nonumber\\
    &+4t{/}\rho\geq\sum\nolimits_{m=1}^M\left(a_m+\delta_{ml}\right)^2+\sum\nolimits_{m=1}^M\left(t-\varepsilon_{ml}\right)^2\nonumber\\
    &+4f_{ml}^{\left(n\right)}\left(t,a_m,\varepsilon_{ml}\right).\label{P5: constraint1}
\end{align}
The communication quality constraint \eqref{P4: constraint2} is reformulated using the same methodology as applied to constraint \eqref{P2:const1}, as
\begin{align}
        &\frac{\left(2N\sum_{m=1}^M\sqrt{a_m\eta_{mk}^c}\gamma_{mk}\right)^2}{\varsigma}+\sum\nolimits_{m=1}^M\left(a_m-\mu_{mk}\right)^2\geq\nonumber\\
        &\sum\nolimits_{m=1}^M\left(a_m+\mu_{mk}\right)^2+4N\sum\nolimits_{m=1}^M\sum\nolimits_{l=1}^L\eta_{ml}^s\beta_{mk}+4{/}\rho.
\end{align}
In contrast, \eqref{ieq: apr1} is applied to obtain a concave lower bound for the left-hand side, yielding the constraint as
\begin{align}
        &\frac{2N\sum\nolimits_{m=1}^M\sqrt{a_m^{\left(n\right)}\left(\eta_{mk}^c\right)^{\left(n\right)}}\gamma_{mk}}{\varsigma}\left(4N\sum\nolimits_{m=1}^M\sqrt{a_m\eta_{mk}^c}\gamma_{mk}\right.\nonumber\\
        &\left.-2N\sum\nolimits_{m=1}^M\sqrt{a_m^{\left(n\right)}\left(\eta_{mk}^c\right)^{\left(n\right)}}\gamma_{mk}\right)+\nonumber\\
        &\sum\nolimits_{m=1}^M\left(a_m^{\left(n\right)}-\mu_{mk}^{\left(n\right)}\right)\left(2\left(a_m-\mu_{mk}\right)-\left(a_m^{\left(n\right)}-\mu_{mk}^{\left(n\right)}\right)\right)\geq\nonumber\\
        &\sum\nolimits_{m=1}^M\left(a_m+\mu_{mk}\right)^2+4N\sum\nolimits_{m=1}^M\sum\nolimits_{l=1}^L\eta_{ml}^s\beta_{mk}+4{/}\rho.\label{P5: constraint2}
\end{align}

Based on the convex approximations, the reformulated optimization problem is presented as
\begin{align}
        \bm{\left(\mathrm{P5}\right):}&\min_{\bm{a},\bm{\eta}^c,\bm{\eta}^s,t}\;t+\lambda\sum\nolimits_{m=1}^Ma_m-a_m^{\left(n\right)}\left(2a_m-a_m^{\left(n\right)}\right)\nonumber\\
        &\mathrm{s.t.}\;\eqref{P5: constraint1}, \eqref{P5: constraint2}, \eqref{P3: constraint3}, \eqref{P2:const4}, \eqref{P2:const5}, \eqref{P2:const6}.\nonumber
\end{align}

Similar to the communication-prioritized optimization problem, this convex program can be efficiently solved using CVX within the \ac{SCA} framework.
%

\section{Numerical Simulations}
\label{Chap4:numerical_simu}


    
    

This section presents numerical results validating the proposed optimization frameworks and evaluating their performance. 
We consider a $500\times500$ $\mathrm{m}^2$ area where $M$ \acp{AP} and $K$ \acp{UE} are uniformly distributed, and $L$ sensing zones are randomly placed. 
Each \ac{AP} is assumed to possess prior knowledge of the angular directions to all sensing zones for target detection. 
%
The system parameters are set to $ (M, N, L) = (32, 8, 2), \, \nu = 0.5, \varsigma = 4$ dB, and $\kappa = 2$ dB.
The network parameters are set to $\tau=200$ and $\tau_t=K+L$. 
The transmit powers for data and pilot signals are set to $\rho=1$ W and $\rho_t=0.25$ W, respectively.

\begin{figure}[!t]
\centering
\includegraphics[width=0.75\columnwidth]{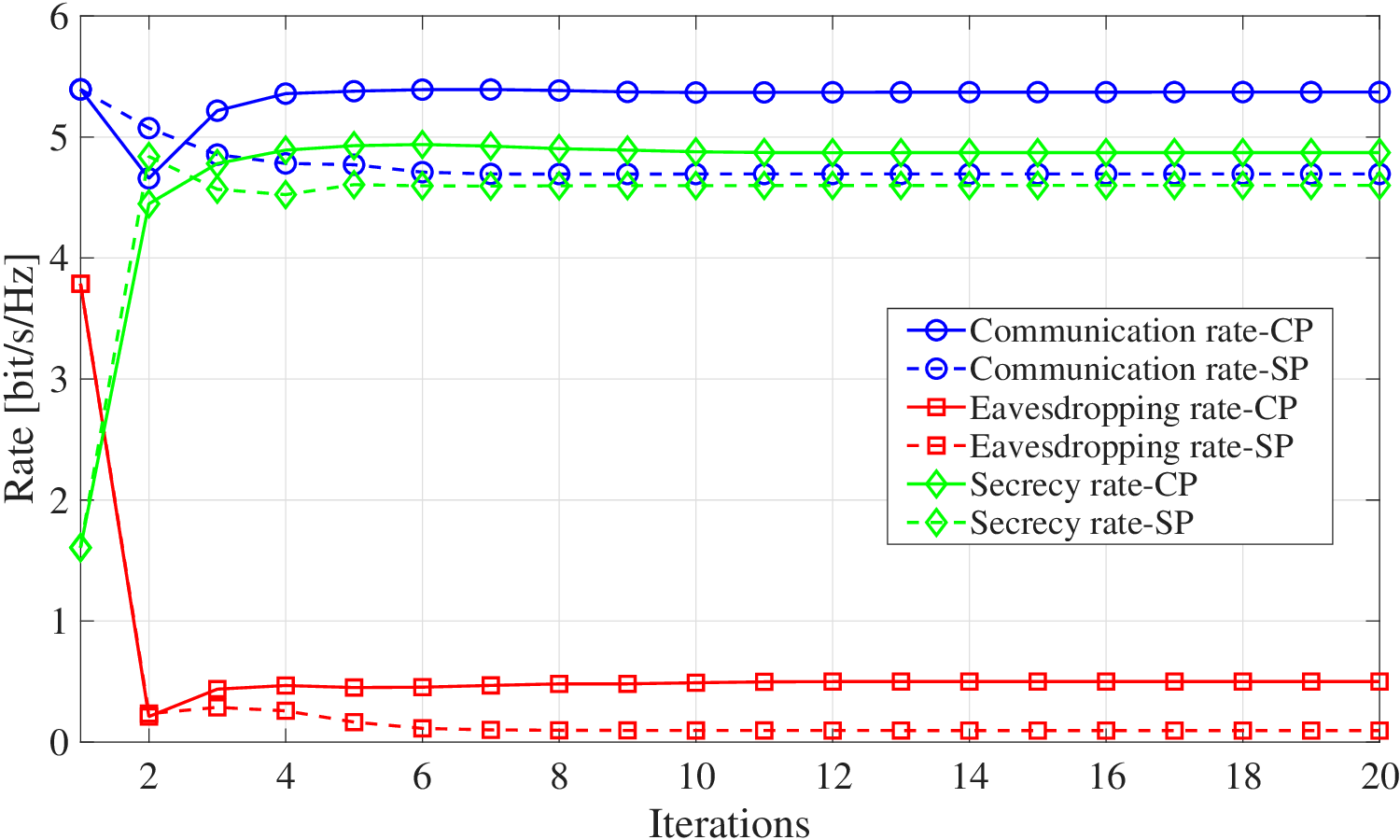}
\vspace{-2mm}
\caption{Convergence of communication, eavesdropping, and secrecy rates.}
\vspace{-1mm}
\label{fig: opt_ite}
\end{figure}

Fig.~\ref{fig: opt_ite} illustrates the trade-off between the two proposed optimization strategies: \ac{CP} and \ac{SP}.
Each strategy excels in its primary objective.
The \ac{SP} strategy lowers the eavesdropping rate by allocating more power to sensing signals, which makes the eavesdropper's channel worse but also lowers the communication rate.
%
%
Conversely, the \ac{CP} strategy maintains a higher communication rate by prioritizing \ac{C-AP} resource allocation, benefiting \acp{UE} but leading to an increased eavesdropping rate due to lower sensing power.
Overall, the \ac{CP} attains a higher secrecy rate, as excessive power allocation to sensing yields diminishing security gains.

\begin{figure}[!t]
\centering
\includegraphics[width=0.75\columnwidth]{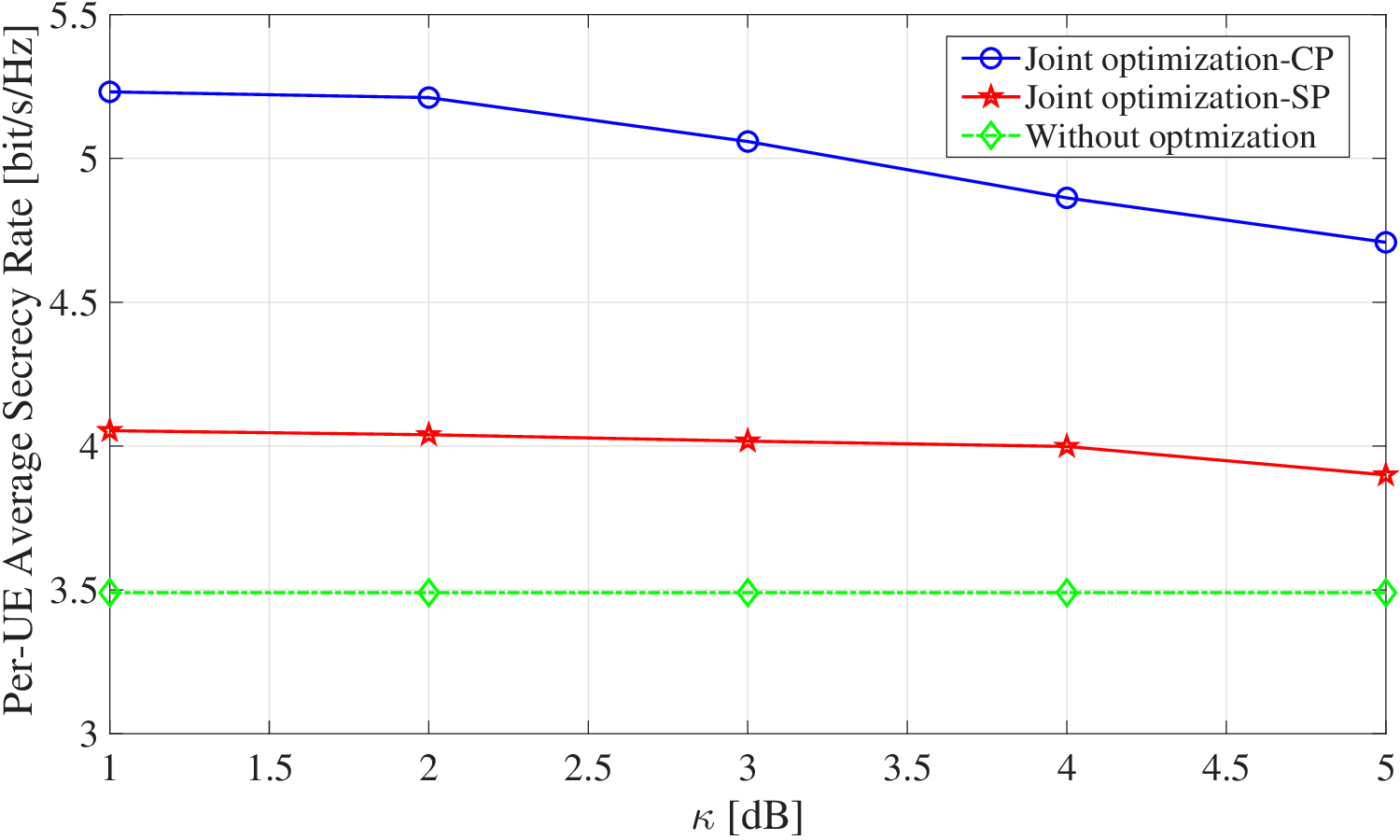}
\vspace{-2mm}
\caption{Comparison of per-\ac{UE} average secrecy rate under different $\kappa$.}
\vspace{-1mm}
\label{fig: opt_kappa}
\end{figure}

\begin{figure}[!t]
\centering
\includegraphics[width=0.75\columnwidth]{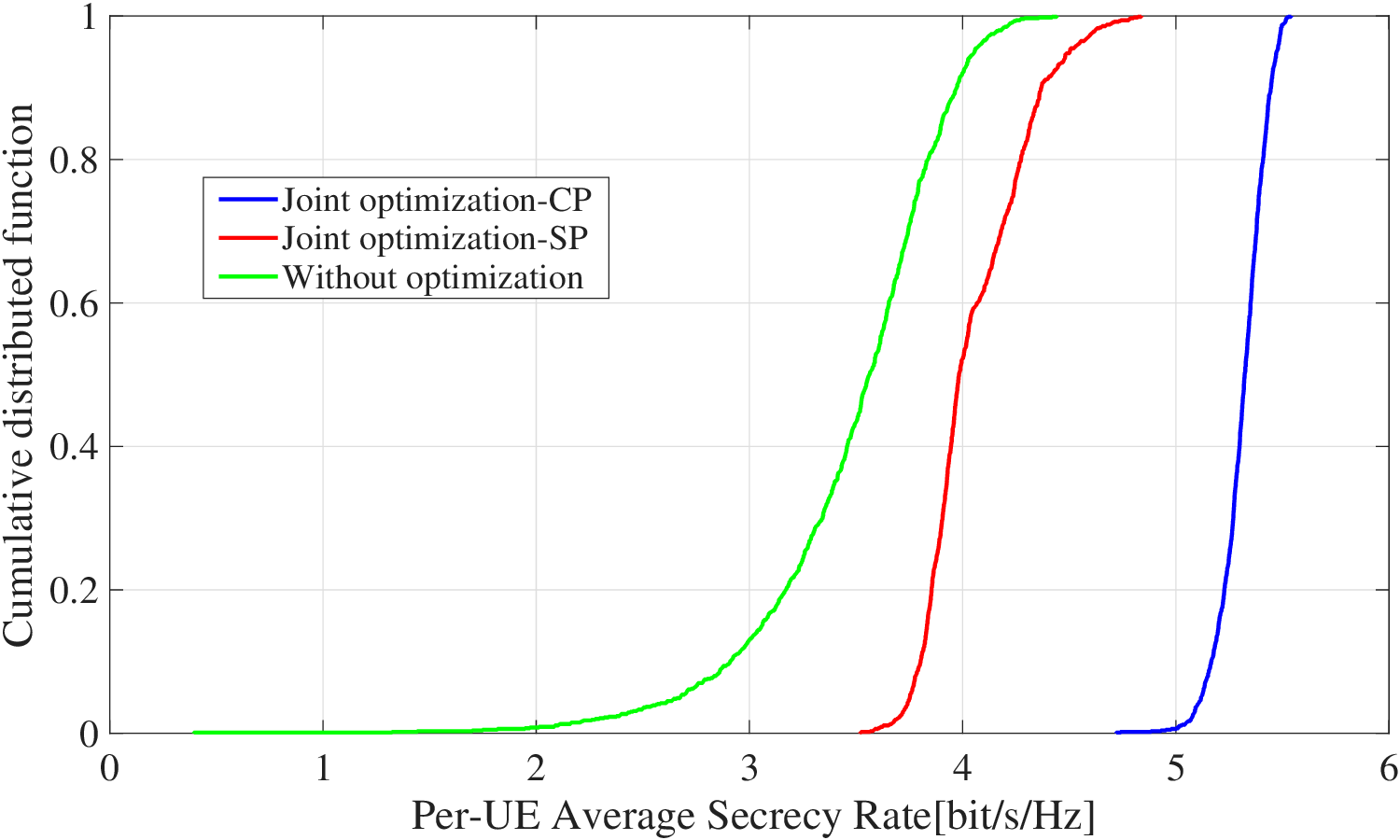}
\vspace{-2mm}
\caption{CDF of the per-\ac{UE} average secrecy rate.}
\vspace{-4mm}
\label{fig: opt_cdf}
\end{figure}

Fig.~\ref{fig: opt_kappa} examines the impact of the minimum sensing requirement $\kappa$ on the secrecy rate performance for both optimization strategies, using an average power allocation scheme used as a baseline.
The \ac{CP} strategy significantly enhances secrecy rates by primarily allocating power to communication links, while ensuring minimal power is dedicated to sensing to fulfill the requirement.
However, as $\kappa$ increases, the \ac{CP} strategy suffers severe performance degradation since more power must be redirected from communication to sensing, reducing user rates and consequently secrecy performance.
%
Conversely, the \ac{SP} strategy maintains robust performance under increasing $\kappa$. 
The inherent allocation of substantial power to sensing facilitates eavesdropper suppression, allowing higher sensing demands to be met without significant power redistribution, thereby ensuring stable secrecy rate performance across the parameter range.

Fig.~\ref{fig: opt_cdf} presents the \ac{CDF} of secrecy rates for both optimization strategies, benchmarked against the average power allocation scheme. 
The \ac{CP} strategy consistently outperforms the others, achieving higher secrecy at all probability levels. 
Specifically, it attains a median secrecy rate of $5.1$ bit/s/Hz, substantially exceeding that of the \ac{SP} method. 
Although the \ac{SP} strategy yields lower secrecy rates than the \ac{CP} strategy, it still provides considerable improvement over the baseline, maintaining secrecy rates above $3.5$ bit/s/Hz and demonstrating robust security performance.

\section{CONCLUSION}
\label{Chap5:conc}


This paper investigated a multi-static \ac{CF-mMIMO} \ac{ISAC} system employing \ac{AP} selection.
Through closed-form derivations of the communication rate, eavesdropping rate, and \ac{MASR}, we quantified the interplay between communication and sensing functions while demonstrating the security benefits of \ac{AP} selection. 
Optimization problems were proposed to maximize communication rates or minimize eavesdropping rates under sensing performance and power budget constraints. 
Numerical results validated the theoretical analysis and confirmed the effectiveness of the proposed power allocation framework.




\section*{Acknowledgement}

This work was supported 
in part by JST, CRONOS, Japan Grant Number JPMJCS24N1, and in part by MIC/FORWARD under Grant JPMI240710001.

\bibliographystyle{REF/IEEEtran}
\bibliography{REF/IEEEabrv,REF/conf_abbrv,REF/ref}

\end{document}